%% file: main.tex
\documentclass[runningheads]{llncs}
\usepackage[T1]{fontenc}
\usepackage{graphicx}
\usepackage{hyperref}
\usepackage{color}

\usepackage{xcolor}
\usepackage{amsmath,amssymb,amsfonts}  %
\usepackage{physics}
\usepackage{booktabs}
\usepackage{multirow, multicol}
\usepackage{subcaption}
\usepackage{tikz}
\usepackage{algorithm}
\usepackage{algpseudocode} 

\algnewcommand\algorithmicinput{\textbf{Input:}}
\algnewcommand\Input{\item[\algorithmicinput]}
\algnewcommand\algorithmicoutput{\textbf{Output:}}
\algnewcommand\Output{\item[\algorithmicoutput]}
\algnewcommand\algorithmicgoto{\textbf{go to}} %
\algnewcommand\Goto{\algorithmicgoto\ }

\newcommand{\pieval}{$\Pi_{\text{eval}}$}
\newcommand{\pisel}{$\Pi_{\text{sel}}$}
\def\barroman#1{\sbox0{#1}\dimen0=\dimexpr\wd0+1pt\relax
  \makebox[\dimen0]{\rlap{\vrule width\dimen0 height 0.06ex depth 0.06ex}%
    \rlap{\vrule width\dimen0 height\dimexpr\ht0+0.03ex\relax 
            depth\dimexpr-\ht0+0.09ex\relax}%
    \kern.5pt#1\kern.5pt}}

\begin{document}
\title{Reusable Test Suites for Reinforcement Learning}
\author{Jørn Eirik Betten\inst{1,3}\orcidID{0009-0009-2201-997X} \and Quentin Mazouni\inst{1}\orcidID{0009-0003-3519-5514} \and Dennis Gross\inst{1}\orcidID{0009-0001-5734-0538} \and Pedro Lind\inst{2,3}\orcidID{0000-0002-8176-666X} \and Helge Spieker\inst{1}\orcidID{0000-0003-2494-4279}}
\authorrunning{J.E. Betten et al.}
\institute{Simula Research Laboratory, Kristian Augusts gate 23, Oslo, Norway \and
School of Economics, Innovation and Technology, Kristiania University of Applied Sciences, Kirkegate 24-26, 0153 Oslo, Norway
\and
Oslo Metropolitan University, Pilestredet 46, Oslo, Norway\\
\email{jorneirik@simula.no}}
\maketitle
\begin{abstract}

\emph{Reinforcement learning (RL)} agents show great promise in solving sequential decision-making tasks. However, validating the reliability and performance of the agent policies' behavior for deployment remains challenging.
Most \emph{reinforcement learning} policy testing methods produce test suites tailored to the agent policy being tested, and their relevance to other policies is unclear.
This work presents \emph{Multi-Policy Test Case Selection (MPTCS)}, a novel automated test suite selection method for RL environments, designed to extract test cases generated by any policy testing framework based on their solvability, diversity, and \emph{general} difficulty. 
MPTCS uses a set of policies to select a diverse collection of \emph{reusable policy-agnostic} test cases that reveal typical flaws in the agents' behavior. The set of policies selects test cases from a candidate pool, which can be generated by any policy testing method, based on a difficulty score. 
We assess the effectiveness of the difficulty score and how the method's effectiveness and cost depend on the number of policies in the set. Additionally, a method for promoting diversity in the test suite, a discretized general test case descriptor surface inspired by quality-diversity algorithms, is examined to determine how it covers the state space and which policies it triggers to produce faulty behaviors. 

\end{abstract}

\section{Introduction}
Autonomous agents trained through the \emph{reinforcement learning (RL)} paradigm~\cite{sutton_reinforcement_2015} have demonstrated impressive capabilities in complex sequential decision-making tasks such as game playing~\cite{silver_mastering_2016}, robotics~\cite{nvidia_nvidia_2025}, and autonomous driving~\cite{waymo_why_2021}.

In RL, agent policies are trained through trial-and-error, given feedback signals as rewards or penalties within an environment. A \emph{neural network} typically represents the policy, mapping environment state observations to action choices. However, these models, sometimes referred to as black-box models, are notoriously opaque~\cite{vouros_explainable_2023,milani_explainable_2023}, and may therefore require extensive analysis to extract rules for how the model responds to the input~\cite{molnar_interpretable_2025}.  Furthermore, the vastness of the state spaces of environments, which confines all possible states the agents can find themselves in, excludes exhaustive testing as a viable option.
Consequently, efficiently assuring the reliability, safety, and performance of RL agents becomes a particularly challenging task. 

To increase our confidence in RL-based automated software, the field of \emph{RL policy testing} has therefore developed methods to test policy behavior~\cite{pang_mdpfuzz_2023,mazouni_testing_2024,zolfagharian_search-based_2023,ul_haq_many-objective_2023,li_generative_2023} as well as hybrid approaches, that either prevent unsafe behaviors~\cite{zolfagharian_smarla_2024,alshiekh_safe_2017} or verify the agent by policy sampling to make the verification tractable~\cite{pranger_test_2024,gross_cool-mc_2022}. RL policy testing methods typically generate \emph{test cases}, e.g., an environment state a policy can act from, designed to expose a flaw in the behavior of a policy.

A common feature of RL policy testing is that the policy under test is involved in the test case generation process.
In particular, the test cases aim to fail specifically the tested policy.
Even though some methods optimize for additional qualities of the resulting test suite, such as diversity~\cite{mazouni_testing_2024,pang_mdpfuzz_2023,li_generative_2023} or robustness~\cite{zolfagharian_search-based_2023,ul_haq_many-objective_2023}, their reusability for other trained policies is unclear. 
Such policy-specific test suites make sense if their purpose is \emph{solely to validate the policy under test}, which is often the case. 
However, the computational load associated with generating test suites for one policy is large with any method, and in practical settings, it is common to train more than one policy, because hyperparameters need to be optimized, and there are performance variations from the stochastic training process. 
Additionally, ensuring that all the generated test cases are \emph{solvable}, which refers to the existence of an action sequence from the environment state that passes the testing criterion, requires expert domain knowledge to design sufficient restrictions to the search space. An unsolvable test case does not give information about the quality of the policy, and policy testing techniques require extra analysis to confirm that the generated test cases are solvable.

We propose to test RL policies by providing \emph{policy-agnostic} test cases, which signifies a shift in the objective of the test suite from revealing faults in the policy under test toward exposing behavioral flaws of any policy. 
To elaborate, the purpose of the policy-agnostic test suite is to present behavioral challenges that expose typical weaknesses in any policy. 
Having a collection of policy-agnostic test cases may expedite RL agent development, because the presence of common weaknesses in the policy behavior may be identified early in the development phase. 
Ordinary behavioral shortcomings of policies then become easier to identify and overcome. Consequently, policies reaching the final testing phases of agent development are already tested for general flaws, reducing the testing (and computational) burden currently resting on policy testing techniques.
For the policy-agnostic test suite to be effective, the test cases must assert the absence of behavioral deficiencies, meaning that the behavioral challenges should be as fault-revealing, or \emph{difficult}, as possible. Additionally, it should maximize the coverage of plausible flaws in behavior, meaning that the policy-agnostic test suite should provide a \emph{diverse} collection of behavior challenges~\cite{ma_diversity-oriented_2025,pang_mdpfuzz_2023,mazouni_testing_2024}.

We present \emph{Multi-Policy Test Case Selection (MPTCS)}, a novel approach for selecting policy-agnostic test cases in RL for any testing criterion, compatible with any RL environment.
The idea behind MPTCS is to leverage a \emph{set} of behaviorally distinct policies to select confirmed solvable but difficult test cases resulting from any test case generation method.
MPTCS aims to build a policy-agnostic test suite that reveals a wide range of common deficiencies in policy behavior by selecting for three desirable attributes of the test cases: 

(1) \emph{Solvability}. Ensuring the test cases are solvable by at least one policy, filtering out inherently unsolvable challenges. 

(2) \emph{Difficulty}. Prioritizing test cases where a significant fraction of the set of policies fails, thus favouring generally challenging test cases. 

(3) \emph{Diversity}. Promoting a varied set of test cases selected by employing a niche-based selection mechanism inspired by quality-diversity algorithms~\cite{mouret_illuminating_2015}, to favor the exposure of different types of policy weaknesses. 
Considering these three attributes, we evaluate three aspects of MPTCS in four RL environments. First, the effectiveness of the difficulty score selection criterion, second, the tradeoff between the extra computational load and the general difficulty of the test cases when varying the number of policies used in MPTCS, and, third, a generally designed archive selection inspired by quality-diversity algorithms~\cite{mouret_illuminating_2015} for enhanced diversity of the test suite. 
Results show that the MPTCS test suites successfully expose more weaknesses in the policies' behaviors than single-policy methods, and that the archive selection can enhance the diversity of the test suite.

The contributions of this paper are: (\barroman{I}) Identifying the usefulness of policy-agnostic test suites, and specifying the problem of \emph{policy-agnostic testing}; 
(\barroman{II}) MPTCS, a method that, given a testing criterion and a set of policies, selects a diverse collection of confirmed solvable but difficult test cases in RL environments; 
(\barroman{III}) A comparison of the general difficulty between test cases selected by a single policy and test cases selected through the difficulty score requiring multiple policies; 
(\barroman{IV}) An investigation into the tradeoff between computational cost and general difficulty when varying the number of policies used to select test cases; 
Finally, (\barroman{V}) an examination of how the generally designed archives inspired by quality-diversity algorithms promote diversity of test suites. 

\section{Background}\label{sec:background}
\subsection{Reinforcement Learning}\label{sec:reinforcement-learning}
We model the environment as a Markov Decision Process (MDP)~\cite{sutton_reinforcement_2015}, which is a tuple $M=\langle\mathcal{S}, \mathcal{A}, R, P, \gamma, \mathcal{S}_{\text{init}}\rangle$, where $\mathcal{S}$ is the state space of the environment, $\mathcal{A}$ is the agent's action space, and $R: \mathcal{S}\times\mathcal{A}\times\mathcal{S}\to\mathbb{R}$ is the reward function, signaling how agent behavior is immediately rewarded for the action it takes in a state. $P:\mathcal{S}\times\mathcal{A}\to\mathcal{S}$ is the transition function of the environment, determining the next state from the previous state and the agent's action, and $\gamma$ is the discount factor, representing the urgency of the agent's accumulation of rewards. The combination of the reward function, the transition function, and the discount factor determines the objective to be maximized, i.e., the goal of the agent. Finally, $\mathcal{S}_{\text{init}}$ defines the set of initial states of a trajectory in the MDP, i.e., the first state at the beginning of each trajectory originates from this set, $s_0 \in S_{\text{init}}$, if not overridden by another state, e.g., for testing purposes.

In RL, policies are trained from interactions between the agent and the environment, aiming to maximize a reward signal. Note that the training process is stochastic, both in what states are encountered by the policies during training and through the initialization of the parameters of the policy models. Trained policies, therefore, typically differ in their behaviors~\cite{fang_provably_2025}, the degree to which depends on the complexity and degrees of freedom in the task's solution space.

\subsection{Software Testing}\label{sec:software-testing-problem}
We represent a formal minimal definition for an instance of the problem of testing software in the form of a tuple, 
    \(P_{ST} = \langle S, D, O, \mathcal{C}, U\rangle\), 
where $S$ represents the \emph{system under test}, and $D$ the \emph{domain} of $S$. 
The \emph{oracle} $O$ is a function, formally, $O(d, S(d))\to\qty{\text{pass}, \text{fail}}$, that determines whether the behavior of $S$ for a given input $d\in D$ is correct or not. The set of constraints under which testing must be performed is represented by $\mathcal{C}$, and typically comprises restrictions on resources spent on testing, such as the computational cost of executing the test suite. Finally, $U$ is the \emph{utility function}, which quantifies the value, effectiveness, or quality of a test suite $T_{\text{sel}}\subseteq D$, and is optimized for. Typical utility functions are coverage metrics, or the number of faults detected.

\emph{The software testing problem}, defined by an instance $P_{ST}=\langle S, D, O, \mathcal{C}, U\rangle$, then becomes the challenge of finding an optimal test suite $T_{\text{sel}}^*$ such that $T_{\text{sel}}^*$ is a subset of the domain, $T_{\text{sel}}^*\subseteq D$,  where all constraints defined in $\mathcal{C}$ are satisified, and that $U(T_{\text{sel}}^*, S, O)$ is optimized.

\subsection{Reinforcement Learning Policy Testing}

In the following, we provide definitions of the test case and test oracle functions used in this work. %

\paragraph{Test Case Definition.}
In this work, we have reduced the complexity of RL testing by making test cases deterministic to allow for a more controlled analysis of our proposed method. That is obtained by defining a test case, $x$, as a tuple of a state of the RL system and the key given to determine all pseudorandomness in the simulation environment's transition function, while stochastic policies are readily transformable into deterministic policies by always choosing the most likely action determined by the policy network. %

Formally, a test case $x$ is a tuple of a simulation environment state, $s_0$, and a key to determine all the pseudo-randomness, which we denote as  $x = \qty(s_0, \text{key})$.

A test case can be executed in the simulation environment following a policy, which creates a trajectory from the deterministic initial state
$s_0$. We denote the execution of the test case, $x=\qty(s_0, \text{key})$, following the policy $\pi$, as 
\begin{equation*}
    \text{exec}\qty(x, \pi) = \langle s_0, a_1,  s_1, r_1, a_2, s_2, r_2, \dots, a_{l-1}, s_l, r_l\rangle = \tau, 
\end{equation*}
where all information about the evolving state of the system, action choices, and rewards is stored in the trajectory, which we denote $\tau$. 

\paragraph{Oracle Definition.} 
The test oracle in policy testing is often based on a testing criterion that defines unwanted behavior, such as, e.g., a violation of a safety criterion, and can be represented by a function that takes in a trajectory and returns a binary pass or fail, as \(TC(\tau) \to \qty{0, 1}\), 
where we assume that $0$ indicates a pass while $1$ indicates a failure. The oracle can then be defined as a composition of the test case execution and the failure criterion, as  
\begin{equation}\label{eq:test-oracle-rl}
    O(x, \pi) = TC(\text{exec}(x,\pi))\to \qty{0, 1}.  
\end{equation}

\section{Related Work}

Our work selects the most generally challenging test cases from a test suite output by any RL policy testing framework.
Policy testing for critical failures has been addressed in several ways, including fuzzing~\cite{pang_mdpfuzz_2023}, search-based testing~\cite{tappler_search-based_2022,biagiola_testing_2024}, reinforcement learning~\cite{ul_haq_many-objective_2023}, model-based~\cite{li_generative_2023}, and metamorphic testing~\cite{eisenhut_automatic_2023}.
Notably, the importance of fault diversity (along with fault maximization) was highlighted by Mazouni et al.~\cite{mazouni_testing_2024}.
They promote fault diversity with quality-diversity optimization, similarly to our work.

In software testing, the technique of \emph{test case prioritization}~\cite{shin_empirical_2019,spieker_reinforcement_2017} or \emph{test suite reduction}~\cite{shi_comparing_2015} is often applied when resources, like time, are scarce, or testing is expensive~\cite{birchler_single_2023}, especially in regression testing settings, or during continuous integration for quick feedback~\cite{li_semantic-aware_2024}. 
Birchler et al.~\cite{birchler_single_2023} use genetic algorithms to prioritize low execution cost and high diversity of test suites executed in expensive self-driving car simulations, to accelerate regression testing of driving software. 
Similarly to our work, the aim is typically to reduce the costs associated with testing key system functionality. 
Our work distinguishes itself from test case prioritization and policy testing in the objective shift of the test case to \emph{general} usability, referring to the policy-agnosticism. As far as we know, applying multiple policies to assess the \emph{general value} of a test is a novel approach.

\section{Problem Definition}
The policy-agnostic testing problem may be posed as a modified software testing problem, where the system is \emph{the set of all possible policies, $\Pi$}, as \(\langle \Pi, D, O, \mathcal{C}, U\rangle\). 
In this case, the domain $D$ is the state space of the RL environment, while the test oracle embodies the function checking the testing criterion. The set of constraints may be determined by the resources available to the testing team. Finally, $U$ promotes the diversity and difficulty of the challenges posed by the test suite toward all policies in $\Pi$. 

Crucially, note that the system under test is the abstraction of \emph{all possible policies} for the RL system. This represents a major shift away from seeking behavioral flaws of the specific policy under test, toward \emph{common behavioral deficiencies}, i.e., policy-agnostic test cases, exhibited by policies in that RL system. 

\section{Multi-Policy Test Case Selection}\label{sec:mptcs}

This section describes our method for selecting policy-agnostic test cases, Multi-Policy Test Case Selection (MPTCS). It begins by presenting a high-level overview of MPTCS. Next, it specifies how MPTCS favors selecting a diverse set of generally difficult and solvable test cases, followed by the implementation procedure. Lastly, it discusses how MPTCS can be integrated with policy testing techniques.

\subsection{Overview of Multi-Policy Test Case Selection}

The Multi-Policy Test Case Selection (MPTCS) method selects test cases based on their difficulty, diversity, and solvability. The difficulty is measured through the number of policies that fail, the solvability by requiring at least one passing policy, and the diversity by organizing the test cases according to the variation in behavior they cause in the policies.

\begin{figure}[t]
    \centering
    \resizebox{\textwidth}{!}{\input{figures/overview2.tikz}}
    \caption{Schematic flow chart of the MPTCS method using an example from the Breakout environment~\cite{young_minatar_2019} with a selection set of five policies. In step 0, we receive a candidate test case, which all policies in the selection set execute (step 1). From the resulting trajectories, we extract the test case descriptors (step 2) and calculate the difficulty score via Equation~\ref{eq:difficulty-score}. In step 3, we replace the cell occupant with the candidate if it has a higher difficulty score than the occupant.}
    \label{fig:overview}
\end{figure}
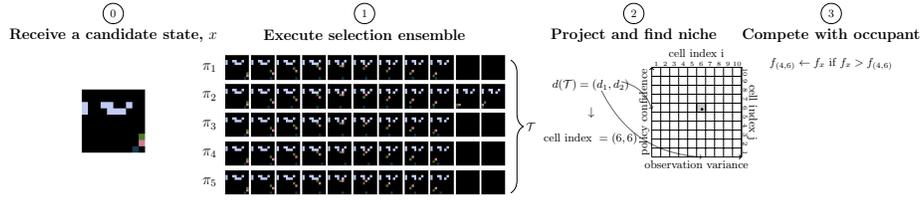

Figure~\ref{fig:overview} shows the workflow of MPTCS with an example from Breakout using five policies as \pisel{}. MPTCS receives a test case candidate state (step 0), all policies in \pisel{} execute it, resulting in $\vert\Pi_{\text{sel}}\vert=5$ trajectories (step 1). The five trajectories are projected onto a descriptor surface and given an archive index (step 2). The archive index determines the test case niche it belongs to, and the test case candidate is included in the archive at the index if the difficulty is higher than the current niche occupant's (step 3).

\subsection{Fostering Diverse, Difficult, and Solvable Test Cases}
This section presents how MPTCS favors diverse, difficult, and solvable test cases in its selection of test cases. We assume a set of $m$ strong policies, $\Pi$, all of which execute a test case candidate, $x$.   

\textbf{Confirmed solvable test cases.}\label{sec:solvable-test-cases}
For a test case to be capable of exposing a flaw in the policy behavior, it needs to be \emph{solvable}. A test case $x$ is solvable if a policy $\pi$ can pass the test oracle, i.e., $O(x,\pi) = 0$. MPTCS only selects confirmed solvable test cases by selecting test case candidates for which at least one policy in a set of policies, $\Pi$, has passed. Formally, we formulate this criterion as $\mathbb{I}\qty(\exists \pi\in\Pi: O(x, \pi) = 0)$, which is the function indicating whether at least one policy in \pisel{} passes the test oracle.

\textbf{Prioritizing difficulty.}
In MPTCS, each test case candidate is scored by the observed failure rate of the selection set of policies. Formally, the candidate difficulty score is given as 
\begin{equation}\label{eq:difficulty-score}
    \text{difficulty}(x, \Pi) = \mathbb{I}\qty(\exists \pi\in\Pi: O(x, \pi) = 0)\frac{1}{m}\sum_{j=1}^m O\qty(x, \pi_j), 
\end{equation}
which results in $\text{difficulty}\qty(x, \Pi)\in\qty{0, \frac{1}{m}, \frac{2}{m}, \dots, \frac{m-1}{m}}$. The change in maximal difficulty score when adding one policy to the selection set becomes $\frac{1}{(m+1)m}$ assuming $m$ policies in the original selection set.

\textbf{Promoting diversity.}
To promote diversity in our test cases, we draw inspiration from quality-diversity algorithms~\cite{mouret_illuminating_2015} that consider both the diversity and quality when selecting test cases. Specifically, we define a two-dimensional descriptor space for all test cases, $\mathcal{D}\in\mathbb{R}^2$, based on two metrics calculated on the trajectories from executing the selection set of policies on a test case candidate.
The descriptor space is discretized into equal-sized two-dimensional cells, which are referred to as \textit{niches}. Executing a test case candidate $x$ with all $m$ policies in the selection set results in $m$ trajectories starting from the same state (see step 1 in Figure~\ref{fig:overview}), which can be projected onto the descriptor space by calculating the metrics (step 2 in Figure~\ref{fig:overview}). The resulting descriptor determines the niche of the test case, and only the test case with the highest difficulty score within the niche, the \emph{elite}, will remain in the test suite (step 3 in Figure~\ref{fig:overview}). We refer to the collection of all niches and their occupants as the archive, which comprises the test suite.

\subsection{Implementation}
Algorithm~\ref{alg:mptcs} presents a general pseudocode for implementing MPTCS. Given a test case candidate input, we execute all policies from it, and use the resulting trajectories to map it onto the corresponding cell in the discretized descriptor space. The current cell occupant is replaced if the candidate test case difficulty score exceeds the current occupant's.

\begin{algorithm}[t] 
\caption{Multi-Policy Test Case Selection (MPTCS)}
\label{alg:mptcs}
\begin{algorithmic}[1] %
    \Input Test case candidate $x$, archive $\mathcal{A}$, and a set of policies, \pisel{}.
    \Output Updated archive $\mathcal{A}$. 
    \Statex %
    \Procedure{MPTCS}{$x, \mathcal{A}, \Pi_{\text{sel}}$}
        \State $\mathcal{T}=\emptyset$ \Comment{Initialize empty set of trajectories.}
        \ForAll{$\pi\in\Pi_{\text{sel}}$} 
            \State $\tau = \text{exec}\qty(x, \pi)$ \Comment{Execute policy from the test case.}
            \State $\mathcal{T}\gets\mathcal{T}\cup\qty{\tau}$ \Comment{Add the trajectory to the set of trajectories.}
        \EndFor
        \State $\mathbf{d}\gets\text{descriptor metrics}\qty(\mathcal{T})$ \Comment{Calculate test case descriptors from $\mathcal{T}$.}
        \State $\text{score}\gets\text{difficulty}(x, \Pi_{\text{sel}})$ \Comment{Calculate the difficulty score.}
        \State $\text{index}\gets\text{cell index}(\mathbf{d})$ \Comment{Find index of archive corresponding to $\mathbf{d}$.}
        \State $\text{occupant}\gets\mathcal{A}\qty[\text{index}]$ \Comment{Extract the occupant of the cell.}
        \If{$\text{occupant.score} < \text{score}$} \Comment{If $x$ is more difficult, add it to archive.}
            \State $\mathcal{A}\qty[\text{index}]\gets\text{new occupant \(x\) with score \(score\)}$
        \EndIf
        \State \textbf{return} $\mathcal{A}$
    \EndProcedure

\end{algorithmic}
\end{algorithm}

\subsection{Integration in Policy Testing}
Prerequisites for MPTCS to be deployable in an RL environment are (1) a test case definition, (2) a way to execute test cases and store the resulting trajectories, (3) a testing criterion, (4) a method for generating test cases, (5) a set of distinct policies, and (6) a defined test case descriptor surface. MPTCS requires multiple policies and a defined test case descriptor surface, but has no additional environment dependencies beyond other policy testing methods. 

We propose two levels of MPTCS integration with existing policy testing techniques. First, as a pure post-generation filter, and second, by applying the difficulty and diversity objectives to the test case search. The pure post-generation filter does not modify the policy testing method, but filters the generated test case candidates through Algorithm~\ref{alg:mptcs} after executing the test cases with \pisel{}. Applying the difficulty and diversity objectives to the test case search objective refers to modifications in the policy testing method that shift the search objective towards the selection objective. E.g., the difficulty score and the archive can, in an evolutionary search, be the fitness function and population, respectively.

\section{Experiments}
We empirically evaluate MPTCS for finding policy-agnostic test suites for RL agents.
First, the section poses the research questions we aim to answer. Next, it gives the experimental setup, describing the test bed of RL environments used, briefly explaining how policies were trained and partitioned into sets, and detailing the test case descriptors and generators used in this work. Finally, the investigations into the research questions are described, with the results subsequently presented.  

\subsection{Research Questions}\label{sec:research-questions}
We evaluate MPTCS on three accounts: its effectiveness in finding difficult test cases, how the number of policies in the selection set affects the tradeoff between computational cost and quality of the resulting test suite, and finally, the effectiveness of the niche selection for diverse test cases. 
These accounts are summarized in the following research questions:  
\begin{itemize}
    \item[\textbf{RQ1}] \emph{Is selection based on the difficulty score (Equation~\ref{eq:difficulty-score}) making the test suite more challenging?}
    \item[\textbf{RQ2}] \emph{What is the tradeoff between computational cost and test case quality when varying the number of policies in the selection set?}
    \item[\textbf{RQ3}] \emph{Does the archive's niche structure increase test suite diversity?}
\end{itemize}

\subsection{Experimental Setup}
\subsubsection{Environments}
We evaluate MPTCS on four compact versions of Atari 2600 games from the MinAtar collection~\cite{young_minatar_2019}: Asterix, Breakout, Seaquest, and Space Invaders.
Specifically, we use the Pgx-implementation of MinAtar~\cite{koyamada_pgx_2023}.

In Asterix, the goal is to collect gold while dodging enemies. In Breakout, the agent's goal is to remove the wall of bricks by deflecting the ball with a paddle. For Seaquest, the agent controls a submarine that needs to save divers to refill the oxygen levels of the submarine, while also dodging fish and enemy submarines. In Space Invaders, the agent controls a spaceship attacked by aliens, where the goal is to shoot the aliens before they storm the spaceship. 
For more information about the environments, see the MinAtar documentation~\cite{young_minatar_2019}. %

\subsubsection{Training and Selection of Policy Sets}\label{sec:policy-selection}
More than 100 PPO~\cite{schulman_proximal_2017} policies were trained for 200 million simulation steps in each environment using the script provided in the example folder in the Pgx GitHub repository\footnote{\url{https://github.com/sotetsuk/pgx/tree/8287acd/examples/minatar-ppo}}. The training resulted in high-performing policies, with expected returns -- estimated through 1000 evaluation episodes -- in the ranges $\qty[60, 75]$ in Asterix, $\qty[350, 750]$ in Breakout, $\qty[140, 165]$ in Seaquest, and $\qty[700, 1000]$ in Space Invaders. When the sets of selection and evaluation policies are picked, we order the policies by the collected rewards from 1000 evaluation episodes. First, we pick the best for the selection set, then the second best for the evaluation set, the third for the selection set, and so forth, alternating until both sets are of the desired size. This way, we ensure a similar performance of the policies in the selection and evaluation sets.

\subsubsection{Test Case Descriptors}
We examine a \emph{general} test case descriptor surface. The surface is the Cartesian product of a metric reflecting the diversity of how \pisel{} solves the test case, and one that reflects how confident the policies of \pisel{} are in the test case state. 
Specifically, we take the average variance between the state observations passed to the policies for the diversity of solution strategies and the average confidence of the policy set, measured inversely through the average entropy of their action distributions, for the confidence judgment. The motivation is to examine an environment-independent and general descriptor surface that may be used in many RL environments. 

\subsubsection{Test Case Candidate Generation}\label{sec:candidate-generation}
The MPTCS method relies on test case candidates to select from. In this work, we demonstrate the compatibility of MPTCS with two test case candidate generation approaches, a mutation-only Genetic Algorithm (GA), and MDPFuzz~\cite{pang_mdpfuzz_2023}.  

\textit{Mutation-only GA:} Genetic algorithms evolve populations through reproductive selection according to some fitness objective, inspired by the theory of evolution. Mutation-only GAs perform only asexual reproduction, meaning that a single offspring descends from a single parent. Genetic algorithms are widely used in search-based software testing and policy testing~\cite{deb_fast_2002,zolfagharian_search-based_2023}.

The method has three main components: the fitness objective, the sampling mechanism for reproduction, and the reproduction mechanism itself. First, the difficulty score (Equation~\ref{eq:difficulty-score}) is used as the fitness objective. Second, the fitness score of the individual linearly determines the probability of sampling it to be a parent. The final component is the set of mutation operators, defined on all fields of the state definition, designed not to alter the state out of its valid bounds. The mutation on a field is performed with a probability of 10\%.

In this GA candidate generation, the population is represented by the diversity archive. We iteratively sample 200 individuals from the archive, mutate them, execute them with \pisel{}, calculate the test case descriptors and difficulty score, before we insert them in the archive. Initially, the archive is empty, and we sample from the initial state distribution, $\mathcal{S}_{\text{init}}$, until the archive contains at least 100 test cases. 

\textit{MDPFuzz candidate generation:} MDPFuzz keeps a corpus of test cases to be executed from, and it evolves the corpus using the same asexual reproduction mechanism as in the GA, but the selection of parent states is different. The selection of parent states for the next generation uses two criteria, which they refer to as the \emph{sensitivity} and \emph{freshness} of the simulation trajectories from the test cases. The sensitivity is calculated as the difference in reward accumulations of the simulation trajectories from the parent and offspring, while the freshness is calculated by keeping a probability model (Gaussian mixture model) over visited parts of the space of possible trajectories, for which  the probability of a new trajectory gives a number on how `fresh' the test case is. All failures are collected over all iterations, on which we apply MPTCS as a filtering step. 

Our implementation of MDPFuzz for candidate generation follows the original code~\cite{pang_mdpfuzz_2023}, and we apply MPTCS as a filter on the generated test cases.  

\subsubsection{Setup} For all test cases, a failure is detected if the policy terminates within the first ten steps of execution.
The training was performed on a NVIDIA DGX-2 with a GPU Tesla V100-SXM3-32GB and the experiments, whose open-source implementation is available\footnote{\url{https://github.com/JornEirikBetten/mptcs}}, on a NVIDIA GeForce RTX 4070 Laptop GPU.

\subsection{\textbf{RQ1}: Difficulty Comparison between Single- and Multi-Policy Test Case Selection}\label{sec:rq1}

\paragraph{Description.} First, we investigate the main purpose of MPTCS, that is, how effectively MPTCS finds generally difficult test cases, compared to a single-policy baseline. 
We compare the selection of test case candidates using the MPTCS multi-policy difficulty score and confirmed solvable failures of a single policy.
We apply MPTCS to filter two candidate pools: one was generated through 1,000 generations of the mutation-only GA approach with a 15-policy \pisel{}, and one from MDPFuzz (run with the best-performing policy), containing at least 20,000 failures.
These two use cases demonstrate MPTCS compatibility with two policy testing methods. Additionally, the two candidate pools are different: the MDPFuzz candidates are generated with no additional goal than to fail a single policy, while the GA candidates are generated by using the 15-policy archive as the population, biasing them toward general difficulty.
We then measure the difficulty of the filtered test suites via the failure rates when executing them with \pieval{}, a set of 20 similarly strong policies.

\paragraph{Results \& discussion.} 
Test suites were selected with the single-policy baseline and MPTCS with 15 policies from the same candidate pools for both candidate generation approaches, and repeated five times. 
The aggregated mean failure rates and fractions of confirmed solvable test cases are tabulated in Table~\ref{tab:minatar_difficulties}.
Both experiments show a higher average failure rate of the test suite selected through the multi-policy criterion than the single-policy baseline in all environments. 
However, the magnitude of the difference in difficulty varies between environments and candidate generation approaches. 
When the test case candidates were generated by the GA, the differences were approximately 23\% (Asterix), 15\% (Breakout), 14\% (Seaquest) and 6\% (Space Invaders). 
With MDPFuzz-generated candidates, we observe failure rates of the multi-policy test suites of approximately 5\%, 17\%, 10\%, and 6\% higher than the single-policy baselines. 

When using MPTCS as a filter, the candidates are generated with the same objectives as the original single-policy test case generation. We observe much lower failure rates on the test suites selected from the failures found through MDPFuzz, except in the Space Invaders environment, which consistently gives high failure rates when evaluated by \pieval{}. Consequently, incorporating the multi-policy archive as the population indicates an inflation of the general difficulty of the single-policy baselines, because the candidates are selected based on a multi-policy objective. Therefore, the single-policy test cases selected in a post-generation filtering step from the MDPFuzz candidates are more representative of the typical general difficulty of test suites from single-policy candidate generation approaches.  

Note also how the fractions of confirmed solvable test cases of all generated candidates vary largely between the environments, ranging from $\sim$90\% (GA candidates in Asterix, and MDPFuzz candidates in Seaquest) to $\sim$40\%. 
Considering the high quality of the policies in \pisel{}, we suspect many of these test cases to be unsolvable, and therefore invalid. 

\begin{table}[t]
    \centering
    \caption{Mean failure rates (MFR, in \%) and confirmed solvable test cases (CSTC, in \%) on single-policy (SP) and multi-policy (MP) test suites selected from our mutation-only GA candidate generation with the 15-policy MPTCS archive as the population and MDPFuzz candidate generation with the MPTCS purely as a filtering method, aggregated over five executions.} 
    \label{tab:minatar_difficulties} 
    \scriptsize
    \begin{tabular}{cccccccccc} 
        \toprule
        \multirow[c]{3}{0.12\linewidth}{\shortstack{Candidate \\ generation}} &  \multirow[c]{3}{*}{Metric} & 
        \multicolumn{2}{c}{Asterix} & \multicolumn{2}{c}{Breakout} & \multicolumn{2}{c}{Seaquest} & \multicolumn{2}{c}{Space Invaders}\\
        \cmidrule(lr){3-10}
        &  & SP & MP & SP & MP & SP & MP & SP & MP \\ 
        \midrule 
        \multirow{2}{*}{GA}& MFR &63.8$\pm$4.7 & 87.0$\pm$2.4 & 44.3$\pm$1.4 & 59.8$\pm$0.8 & 77.4$\pm$3.3 & 91.7$\pm$1.3 & 87.9$\pm$0.2 & 93.3$\pm$0.2 \\ 
        & CSTC & 90.5$\pm$6.1 & 100 & 73.1$\pm$3.6 & 100 & 80.2$\pm$10.9 & 100 & 49.0$\pm$1.0 & 100 \\
        \midrule 
        \multirow{2}{*}{MDPFuzz}& MFR & 42.1$\pm$1.7 & 48.3$\pm$1.4 & 23.5$\pm$0.7 & 40.0$\pm$0.8 & 49.8$\pm$5.1 & 60.7$\pm$6.0 & 87.7$\pm$0.3 & 93.8$\pm$0.2 \\
        & CSTC & 74.3$\pm$11.0 & 100 & 62.6$\pm$2.8 & 100 & 89.0$\pm$5.9 & 100 & 38.3$\pm$0.6 & 100 \\
        \bottomrule
    \end{tabular}
    
\end{table}

\subsection{\textbf{RQ2}: Cost-Difficulty Trade-off}\label{sec:rq2}

\paragraph{Description.} In this experiment, we vary the number of policies in the selection set, \pisel{}, to investigate how it affects the difficulty of the resulting test suite. 
We perform GA candidate generation for 1000 iterations from the same initial states using the same archives for both sampling and collection. 
We set up one archive that accounts for a single-policy counterpart of MPTCS, identical to the implementation in Experiment 1 (Section~\ref{sec:rq1}), except that in every iteration of the GA, the candidates are generated from the single-policy archive. The others account for implementations of MPTCS with varying amounts of policies, $\vert\Pi_{\text{sel}}\vert\in\qty{2, 3, 4, 5, 10, 15, 20}$. %
As before, we evaluate the test suites with the failure rates of the independent evaluation set of policies \pieval{}, which is the same independent of $m$.

\begin{figure}[t]
    \centering
    \includegraphics[width=1.0\linewidth]{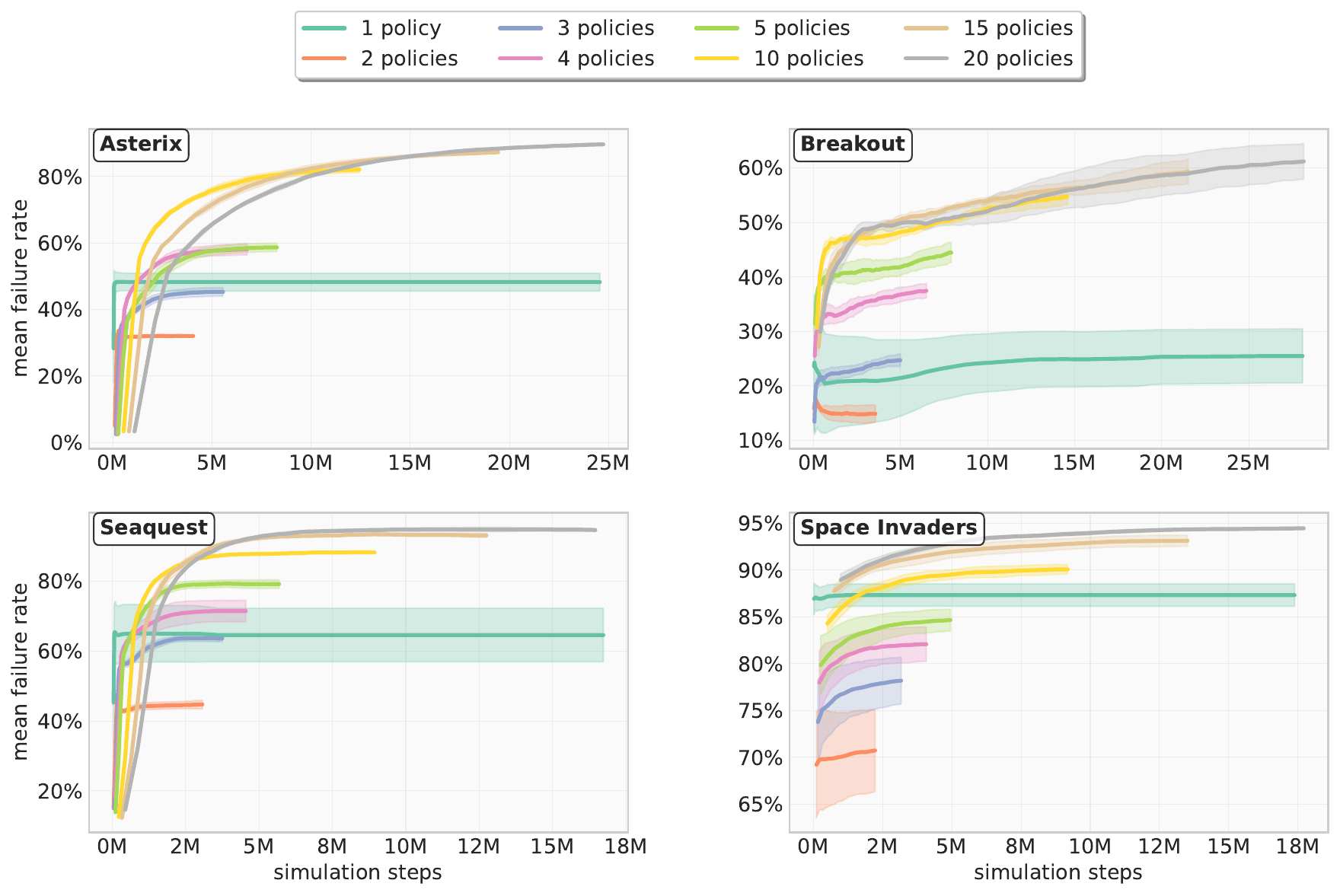}
    \caption{Evolution of the mean failure rates of \pieval{} on the test suites over simulation steps. The results for the single-policy baselines and MPTCS with 2-5, 10, 15, and 20 policies in the selection set are shown for all environments in the test bed. Increasing the number of policies in \pisel{} grows the computational demand linearly and the mean failure rate proportionally to the maximal difficulty score.}
    \label{fig:cost-quality}
\end{figure}

\paragraph{Results \& discussion.} 
Figure~\ref{fig:cost-quality} shows the evolution of the mean failure rate of the test suites over the simulation steps for the single-policy and MPTCS with the different selection set sizes.

For all environments, increasing the number of policies in \pisel{} makes the test cases more difficult. Particularly, we observe that the final average failure rates of the test suites grow diminishingly with the number of policies, reflecting the diminishing increase in the maximal difficulty score. 
One exception is the special case of a single policy, where the selection criterion is modified to collect failures of the single policy, which can be confirmed solvable by any of the 19 remaining policies.
The difficulty score used for the MPTCS implementations gives a non-zero score only when at least one of the policies in \pisel{} passes the test case. Consequently, the single policy baseline has an advantage, because the MPTCS relies on $\vert\Pi_{\text{sel}}\vert$ policies, the ones available in the selection set.  Therefore, when $\vert\Pi_{\text{sel}}\vert=2$, the difficulty score picks test cases where one of the policies in \pisel{} fails and the other passes, it does not necessarily select more difficult test cases than collecting failures of the first policy. However, test cases generated with a true single-policy approach are not confirmed to be solvable here.

Remembering that the single-policy test suites use the single-policy archive as the population in the GA, we observe convergence to lower average failure rates for the single-policy test suites in Figure~\ref{fig:cost-quality} compared to the GA results in Table~\ref{tab:minatar_difficulties}. 
Generally, the single-policy average failure rate reflects the general difficulty of the generated failures for that policy, and we see that this varies greatly between the four environments tested. In particular, the single-policy failures generated in the Space Invaders environment are, on average, failing about 88\% of the policies in \pieval{}, while failures of the single policy are failing only about 25\% of \pieval{} in Breakout.

\subsection{\textbf{RQ3}: Test Suite Diversity}\label{sec:rq3}

\paragraph{Description.} Lastly, we examine how the niche structure promotes test suite diversity.
We compare MPTCS with $\vert\Pi_{\text{sel}}\vert = 15$ to a top-2500 approach, where we collect the 2500 fittest candidates, i.e., those with the highest difficulty score without considering any niches. 
To measure the relative diversity in terms of state space covered by the resulting test suites, we calculate the number of unique states visited from the test cases. Additionally, we measure the relative diversity by examining the distribution of failures among the policies in \pieval{}. %

\paragraph{Results \& discussion.} Unique observations encountered per test case during the evaluation executions of the MPTCS and Top-2500 test suites are shown for each environment in Table~\ref{tab:minatar_diversity}. In Asterix, Seaquest, and Space Invaders, the MPTCS test cases cover more state space than the Top-2500 test cases. However, in the Breakout environment, we see that the test suites' state space coverage is similar, although the Top-2500 is slightly higher. This suggests that the general descriptor metrics (observation variance between policies, and policy confidence of initial state) may not give a test case surface that describes the diversity of test cases well in that specific environment, or that the Top-2500 selection is already selecting very diverse test cases. Breakout is the simplest of the environments, with only three actions and no randomness in the transition function, which may limit the maximal diversity achievable. 
There is a substantial coverage increase in Space Invaders, where the Top-2500 test cases only cover approximately 2.3 unique observations per test case, while the MPTCS archive covers 30.7 unique observations per test case, corresponding to a $\sim$1300\% state coverage increase, while the remaining environments has an increased coverage of about $\sim$50\%.  

Finally, to measure how the test cases spread their challenge across the policies in \pieval{}, we count the frequency of test cases passed by each policy, and normalize it to a distribution over the policies. 
We use the entropy of the distribution of passed test cases among \pieval{}, denoted $\Pi_{\text{eval}}^{\text{pass}}$, as the measure of spread. 
The highest entropy achievable happens when the passes are uniformly distributed, $\log(20)\approx 3.0$. 
We observe that the MPTCS test cases are passed more uniformly by the policies in \pieval{} than the Top-2500 test cases for all environments. In Asterix, the Top-2500 test suite has a 95\% failure rate on all the policies, where one policy is responsible for 91\% of the passed test cases, whereas the passes are almost uniformly distributed for the MPTCS test cases. Both archives distribute passed test cases evenly among \pieval{} in Breakout, suggesting that both test suites are diverse. 
For Seaquest, a few policies are disproportionately passing many test cases for both approaches, most evident for the Top-2500 test suite. These results consolidate the state coverage trend from Table~\ref{tab:minatar_diversity} by further supporting the claim that the MPTCS archive enhances the diversity of the test cases in the test suites for the Asterix, Seaquest, and Space Invaders environments. 

\begin{table}[t]
    \centering
    \caption{Mean failure rates, unique encountered observations per test case, and the entropies of the distributions of passing-frequencies of the policies in \pieval{} after executing \pieval{} on the Top-2500 and MPTCS test suites, aggregated over 5 executions.} 
    \label{tab:minatar_diversity} 
    \scriptsize
    \begin{tabular}{lcc|cc|cc|cc}
        \toprule
        Environment & \multicolumn{2}{c|}{Asterix} & \multicolumn{2}{c|}{Breakout} & \multicolumn{2}{c|}{Seaquest} & \multicolumn{2}{c}{Space Invaders}\\
        \cmidrule(lr){2-3}\cmidrule(lr){4-5}\cmidrule(lr){6-7}\cmidrule(lr){8-9}
        Test case selection & Top-k & MPTCS & Top-k & MPTCS & Top-k & MPTCS & Top-k & MPTCS \\ 
        \midrule 
        Mean failure rate (\%) & 95.0$\pm$0.0 & 92.8$\pm$0.1 & 95.0$\pm$0.0 & 64.4$\pm$0.4 & 95.0$\pm$0.0 & 94.8$\pm$0.0 & 95.0$\pm$0.0 & 94.9$\pm$0.0 \\ 
        Unique observations & 10.6$\pm$0.3 & 15.0$\pm$0.3 & 14.4$\pm$0.0  & 12.2$\pm$0.0 & 11.3$\pm$0.7 & 16.2$\pm$0.3 & 2.3$\pm$0.2 & 30.9$\pm$0.0 \\
        Entropy $\Pi_{\text{eval}}^{\text{pass}}$ & 0.4$\pm$0.0 & 2.8$\pm$0.0 & 2.8$\pm$0.1  & 3.0$\pm$0.0 & 1.1$\pm$0.1 & 1.6$\pm$0.2 & 1.7$\pm$0.2 & 2.9$\pm$0.0 \\
        \bottomrule
    \end{tabular}
\end{table}

\subsection{Threats to Validity}

The selection of environments may inherently bias our experiments and analysis.
To ensure generality, we select four diverse off-the-shelf environments that cover different tasks, e.g., survival or navigation.

Similarly, the available policies might affect the results, as they could have identical behavior or poor performance.
As described in Section~\ref{sec:policy-selection}, we carefully trained a large set of policies to identify the strongest for our experiments, which we divide between \pisel{} and \pieval{}. Still, there are variations in performance that show the differences in policy behavior.

The full testing pipeline with MPTCS combines multiple internal and external components that can influence the final results. We design our experiments to understand these influences by using multiple test case generators, including a baseline from the literature, varying the number of policies used, and removing the test suite archive.

\section{Conclusions}
Policy-agnostic testing reveals typical weaknesses of policies in the RL environment, and builds \emph{reusable} test suites. 
In this work, we introduced a multi-policy method for selecting generally difficult but solvable test cases in policy-agnostic testing, MPTCS.
MPTCS is compatible with most test case generation strategies and can be used to generalize and reduce test suites.

We assessed the effectiveness of MPTCS to find diverse, yet difficult test cases, and investigated the trade-off between simulation costs and test case difficulty when varying the number of policies used to select.
To that aim, we conducted several experiments based on four environments, using high-performing policies and two candidate test case generation methods.
We found that MPTCS successfully increases the difficulty of the test cases compared to standard test case generation, and observed large variations in the policy-agnosticity of test cases selected by a single policy.
Likewise, we measured significant improvements in the test suite diversity over the baseline in terms of environment state coverage and spread of failures in the evaluation policies.
As for the trade-off between cost and test case difficulty, our study revealed that while the former linearly follows the number of selection policies, the latter increases diminishingly, suggesting that MPTCS does not require an extensive set of policies.

In future work, we aim to improve the practical applicability and availability of MPTCS and policy-agnostic test suites.
If test suites and sets of distinct, high-performing policies are made readily available to RL developers, the community of RL developers can better assess policies during development. 

\begin{credits}
\subsubsection{\ackname} 
This work supported by the Norwegian Ministry of Education and Research, the Research Council of Norway (RCN, grant no. 324674 - AutoCSP, 270053 - eX3) and the European Union (grant no. 101091783 - MARS). 

\subsubsection{\discintname}
The authors have no competing interests to declare.
\end{credits}
\bibliographystyle{splncs04}
\bibliography{refs}
\end{document}

%% file: figures/overview2.tikz
\usetikzlibrary{positioning}
\usetikzlibrary{math}
\usetikzlibrary{decorations.pathreplacing}

\begin{tikzpicture}[trim left={(-5.5,0)}]
    \node[circle, draw, draw opacity=1.0] (step_1) at (-5-1, 11) {0}; 
    \node[anchor=center] at (-5-1, 10.4) {\large\textbf{Receive a candidate state,} $x$}; 
    
    \node[inner sep=0pt, anchor=center] (test_case) at (-5-1, 8) {\includegraphics[width=.15\textwidth]{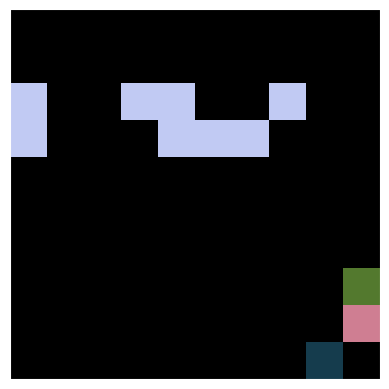}}; 

    \node[circle, draw, draw opacity=1.0] (step_2) at (2-1, 11) {1}; 
    \node[anchor=center] at (2-1, 10.4) {\large\textbf{Execute selection ensemble}}; 

    \node[inner sep=0pt, anchor=center] (p1) at (2-1, 9.5) {\includegraphics[width=.65\textwidth]{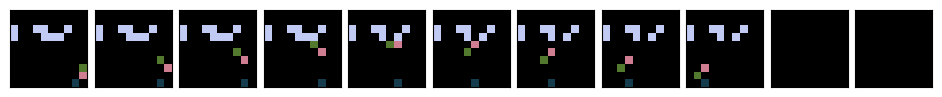}};
    
    \node[inner sep=0pt, anchor=center] (p2) at (2-1, 8.7) {\includegraphics[width=.65\textwidth]{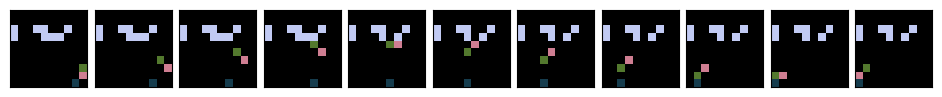}};
    \node[inner sep=0pt, anchor=center] (p3) at (2-1, 7.9) {\includegraphics[width=.65\textwidth]{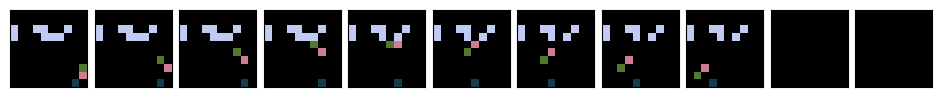}};
    \node[inner sep=0pt, anchor=center] (p4) at (2-1, 7.1) {\includegraphics[width=.65\textwidth]{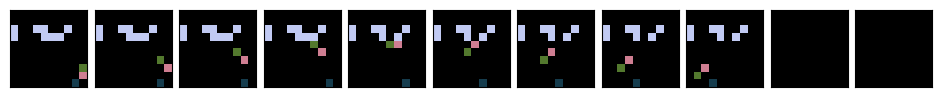}};
    \node[inner sep=0pt, anchor=center] (p5) at (2-1, 6.3) {\includegraphics[width=.65\textwidth]{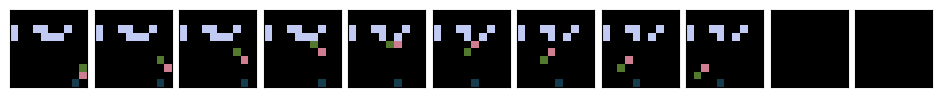}};
    
    \node[anchor=east] at (-2-1, 9.5) {\large{$\pi_1$}}; 
    \node[anchor=east] at (-2-1, 8.7) {\large{$\pi_2$}}; 
    \node[anchor=east] at (-2-1, 7.9) {\large{$\pi_3$}}; 
    \node[anchor=east] at (-2-1, 7.1) {\large{$\pi_4$}}; 
    \node[anchor=east] at (-2-1, 6.3) {\large{$\pi_5$}}; 

    \draw [thick, decorate,decoration={brace,amplitude=10pt},xshift=2pt] (6.0-1,9.7) -- (6.0-1,6.0) node [black,midway,xshift=16pt] {$\mathcal{T}$};   
    
    \node[circle, draw, draw opacity=1.0] at (9-0.5, 11) {2}; 
    \node[anchor=center] at (9-0.5, 10.4) {\large\textbf{Project and find niche}}; 

    \draw[->] (14-5, 6+1) to (14-2.5, 6+1); 
    \node[rotate=0, anchor=center] at (14-3.75, 6+0.8) {observation variance}; 
    \draw[->] (14-5, 6+1) to (14-5,6+3.5);
    \node[rotate=90, anchor=center] at (14-5.2, 6+2.25) {policy confidence}; 
    \draw[-] (14-2.5, 6+1) to (14-2.5, 6+3.5); 
    \node[rotate=0, anchor=center] at (14-3.75, 6+3.9) {cell index i}; 
    \draw[-] (14-2.5, 6+3.5) to (14-2.5, 6+3.5); 
    \node[rotate=-90, anchor=center] at (14-2.2, 6+2.25) {cell index j}; 
    \draw[step=0.25, very thin] (14-5, 6+1) grid (14-2.5, 6+3.5); 

    \foreach \i in {1, 2, ..., 10}
        \tikzmath{ 
            coordinate \x;
            \x = ({14-5.125 + \i*0.25}, {6+3.6}); 
        }
        \node[anchor=center] at (\x) {\tiny{\i}};  

    \foreach \j in {1, 2, ..., 10} 
        \tikzmath{
            coordinate \x; 
            \x = ({14-2.4}, {6+0.875 + \j*0.25}); 
        }
        \node[anchor=center, rotate=-90] at (\x) {\tiny{\j}}; 

    \node[circle, draw, fill=black, inner sep=0.5pt, minimum size=0.1pt] at (14-3.6, 6+2.33) {}; 
    \draw[fill, fill opacity = 0.2] (14-3.75, 6+2.25) rectangle (14-3.5, 6+2.5); 

    \node[anchor=center] at (14-6.7, 6+3) {$d(\mathcal{T}) = (d_1, d_2)$};
    \node[anchor=center, rotate=270] at (14-6.7, 6+2.25) {$\to$}; 
    \node[anchor=center] at (14-6.7, 6+1.5) {$\text{cell index }= (6, 6)$}; 
    \draw[->, draw opacity=0.6] (14-6.4, 6+2.85) to[bend right] (14-3.6, 6+1.0); 
    \draw[->, draw opacity=0.6] (14-5.85, 6+3.1) to[bend left] (14-5, 6+2.33); 

    \node[circle, draw, draw opacity=1.0] at (14, 11) {3}; 
    \node[anchor=center] at (14, 10.4) {\large\textbf{Compete with occupant}}; 

    \node[anchor=center] at (14, 9.6) {$f_{(4, 6)} \gets f_x$ if $f_x>f_{(4, 6)}$}; 
    
\end{tikzpicture}